\documentclass{ws-p8-50x6-00}

\def\dd{\partial}

\begin{document}

\title{Solution of the Boltzmann equation for gluons after a heavy ion
collision\footnote{To appear in the proceedings for ``Strong and
Electroweak matter'' (SEWM2000), Marseille, France, 14-17 June 2000.}}

\author{Jefferson Bjoraker}

\address{University of Minnesota, School of Physics and Astronomy
 Minneapolis, MN. U.S.A.}

\author{ Raju Venugopalan}

\address{Physics Department, Brookhaven National Laboratory
 Upton, NY. 11973, U.S.A.}  


\maketitle

%{\small \noindent  \hfill NUC-MINN-00/24--T}

%{\small  \hfill BNL-NT-00/23}

\abstracts{A non-linear Boltzmann equation\cite{MUELLER} describing the time evolution 
of a partonic system in the central rapidity region after a heavy ion collision is solved
numerically\cite{JEFF}. A particular model of the collinear logarithmic divergences due 
to small angle scattering\cite{ADRIAN,BIRO} is employed in the numerical
solution.  The system is followed until it
reaches kinetic equilibrium where the equilibration
time, temperature and chemical potential are determined for both RHIC
and LHC. }

\section{Introduction}

Heavy ion collisions are already taking place at the Relativistic
Heavy Ion Collider (RHIC) and are planned to take place in a few years at the 
Large Hadron Collider (LHC). The question remains 
whether the hot and dense matter formed after the collision equilibrates to form 
the quark gluon plasma. 
After a heavy ion collision, an initial
distribution of partons, way out of equilibrium\cite{ALEX_RAJU}, 
is formed and strongly influences the subsequent interactions of the
partons with each other which drive the partonic system towards
equilibrium. Whether or not the system equilibrates depends on
if the time it takes to reach equilibrium is shorter than the
hydrodynamic expansion time.

The small $x$ Fock states in nuclei
responsible for multi--particle production at central rapidities
have distributions that are described by a classical effective
field theory (EFT)~\cite{MV}. The classical distributions in a single nucleus 
can be solved analytically\cite{JKMW,Kovchegov}. The classical gluon distribution falls off 
as $1/k_t^2$ at large transverse momentum $k_t$ but saturates at smaller 
$k_t$; growing logarithmically with $k_t$. Non--perturbative, numerical
solutions of the Yang--Mills equations have determined exactly
initial number and energy distribution of gluons after a
collision\cite{ALEX_RAJU} and provide an initial condition for the single particle
gluons distributions in a transport equation. For simplicity, an idealized approximation
for the gluon multiplicity\cite{MUELLER} is used.
Long ago the parton evolution was described by a Boltzmann equation
under the assumption of boost invariance and solved in the relaxation
time approximation\cite{BAYM}. Recently, a non-linear Boltzmann
description of a partonic system, was formulated\cite{MUELLER} and
solved numerically\cite{JEFF}, with no approximation, for a
particular model of the collinear logarithmic divergences due to small
angle scattering\cite{ADRIAN,BIRO}. The time and temperature for the
system to reach kinetic equilibrium for both RHIC and LHC were calculated.
This Boltzmann description takes into account number conserving 
$2\longrightarrow 2$ processes.  $2\longrightarrow 3$ processes, which
may have a large effect\cite{MUELLER2}, were not
considered. 

The distribution of gluons at the very early times after a high energy
heavy ion collision is described by the bulk scale $Q_s$ of gluon
saturation in the nuclear wavefunction. The scale $Q_s$ is the only
scale in the problem, and all observables calculated depend sensitively on it. The
gluon system is followed until it reaches kinetic equilibrium where the equilibration
time, temperature and chemical potential are determined
In this paper we take $\alpha_S = 0.3$ and $N_c=3$.

\section{The Boltzmann equation for gluons}

The Boltzmann equation is:

\begin{equation}
\frac{\dd f}{\dd t}+\vec{v}\cdot \vec{\nabla}f = C(f),
\label{BE1}
\end{equation}

where $C(f)$ is the collision integral and $f = f(t,x,p)$ is the
single particle gluon distribution. Following Baym \cite{BAYM} we
assume that the transverse dimension of the collision volume for
central collision is large enough for $f$ to depend spatially only on the 
coordinate along the collision axis, $z$. Next we assume that the
central rapidity region is invariant under Lorenzt boosts
\cite{BJORKEN}. Under these assumptions we find that 
$\vec{v}\cdot \vec{\nabla}f= v_z\frac{\dd f}{\dd z} =
-\frac{p_z}{t}\frac{\dd f}{\dd p_z}$ \cite{BAYM} where $f = f(t,\vec{p})$ since we
can always choose $z=0$ due to the assumption of boost invariance. 
The l.h.s of eq. (\ref{BE1}) changes accordingly. The collision
integral $C(f)$,  was determined by Mueller\cite{MUELLER}.  The Boltzmann equation 
becomes \cite{MUELLER,JEFF}: 

\begin{equation}
\frac{\dd f(t,p)}{\dd t}-\frac{p_z}{t}\frac{\dd f(t,p)}{\dd p_z} =
\lambda n(t)L\nabla^2_{\vec{p}}\left(f\right)+2\lambda
n_{-1}(t)L\nabla_{\vec{p}}\cdot \left(\vec{v}f\right),
\label{BE_FINAL}
\end{equation}

where,

\begin{equation}
n(t) = g_G\int \frac{d^3p}{(2\pi)^3}f ~,~~~
n_{-1}(t) = g_G\int \frac{d^3p}{(2\pi)^3} \frac{f}{p} 
\label{N_DIST}
\end{equation}

where $\lambda = 2\pi\alpha_S^2\frac{N_c^2}{N_c^2-1}$ , $\vec{v} = 
\vec{p}/|\vec{p}|$, $g_G = 2(N_c^2-1)$, $n$ is the number density of
gluons. $\nabla_{\vec{p}}$ refers to differentiation with respect to $\vec{p}$.
Integration of (\ref{BE_FINAL}) with respect to
$d^3\vec{p}$, leads us to the condition $tn\sim constant$. Furthermore,
$L$ represents the logarithmic collinear 
divergence arising from small angle scattering \cite{LIFSHITZ,JEFF}:

\begin{equation}
L = \frac{1}{2}\log\left(\frac{\langle p \rangle^2}{m_D^2}\right),
\label{DIV}
\end{equation}

where $\langle p \rangle$ is the average energy and $m_D$ is the Debye
screening mass defined as ~\cite{ADRIAN}:

\be
m_D^2 = \frac{\alpha_S N_c}{\pi^2}\,\int {d^3 p\over |\vec{p}|}\, f(p) \, .
\label{Debye1}
\ee

\section{Initial conditions}

We solve eq. (\ref{BE_FINAL}), using the numerical method discussed
in a previous paper\cite{JEFF}. The initial gluon multiplicity,
$dN/d^2p_{\bot}$, for high energy nuclear collisions can be calculated 
along with the formation time of the gluons that are produced in the collision.
Therefore the initial gluon single particle distributions $f(t,\vec{p})$
required to solve eq. (\ref{BE_FINAL}) are
determined:

\begin{equation}
f(t,\vec{p}) = 
\frac{(2\pi)^3}{g_G}\frac{dN}{d^2p_{\bot}}\frac{\delta (p_z)}{t_0},
\label{F_INIT}
\end{equation}

where $t_0$ is $1.35~ {\rm GeV}^{-1}$  for RHIC and $0.65~ {\rm GeV}^{-1}$ for LHC.
At RHIC $Q_s\sim 1$ GeV and at LHC $Q_s\sim 2-3$
GeV. Although the initial gluon single particle distributions, Eq. (\ref{F_INIT}), were recently 
determined by Krasnitz and Venugopalan\cite{ALEX_RAJU} for simplicity we approximate the $p_{\bot}$
distribution with a step function\cite{MUELLER}:

\begin{equation}
f(t,\vec{p}) =  \frac{c}{\alpha_S N_c t_0}\Theta (Q_s^2-p_{\bot}^2) \delta (p_z).
\end{equation}

\section{Bulk quantities at equilibrium}

Having numerically solved eq. (\ref{BE_FINAL}) for $f$ as a function
of $t$ and $\vec{p}$ it is useful to define bulk quantities of the
gluon system, such as the local energy density $\epsilon$ and the longitudinal
pressure $P_L$,
\be
\epsilon(t) = g_G\int \frac{d^3p}{(2\pi)^3}\,p\,f(t,\vec{p}) \, ~~~,~~~
P_L(t) =g_G\int \frac{d^3p}{(2\pi)^3}\, {p_z^2\over p}\, f(t,\vec{p}) \, .
\label{EP}
\ee

We note also that the transverse pressure  $P_T = \frac{1}{2}(\epsilon-P_L)$.
Furthermore, for Boltzmann statistics, the definition of the entropy density per
particle is:

\begin{equation}
s = -\frac{g_G}{n(t)}\int
f(t,\vec{p})\log f(t,\vec{p}).
\label{ENTROPY}
\end{equation}

At kinetic equilibrium $3P_L = 3P_T = \epsilon$. Therefore from
Bjorken hydrodynamics  \cite{BJORKEN} the energy density $\epsilon
\sim t^{-4/3}$. At kinetic equilibrium, the single particle
distribution is $f(t,p) = \exp\left(\beta(\mu-\mid\vec{p}\mid )\right)$ where
$\beta = 1/T$. Some useful quantities at kinetic equilibrium are:

\begin{equation}
n(t) =\frac{g_GT^3}{\pi^2}\exp\frac{\mu}{T}~,~~~
n_{-1}(t) =\frac{g_GT^2}{2\pi^2}\exp\frac{\mu}{T}~,~~~
\epsilon =3\frac{g_GT^4}{\pi^2}\exp\frac{\mu}{T}.
\label{EQUIL}
\end{equation}

From eq.'s (\ref{EQUIL}), we
note that at kinetic equilibrium (since the number density $n\sim
1/t$) $T\sim t^{-1/3}$, and that $\mu/T\sim constant$ with respect to time.
Furthermore, the temperature at kinetic equilibrium is proportional to the energy
density per particle, $T \sim \epsilon/n$.

\section{Numerical results}

\begin{figure}[tbh]
\centering \leavevmode
\psfig{file=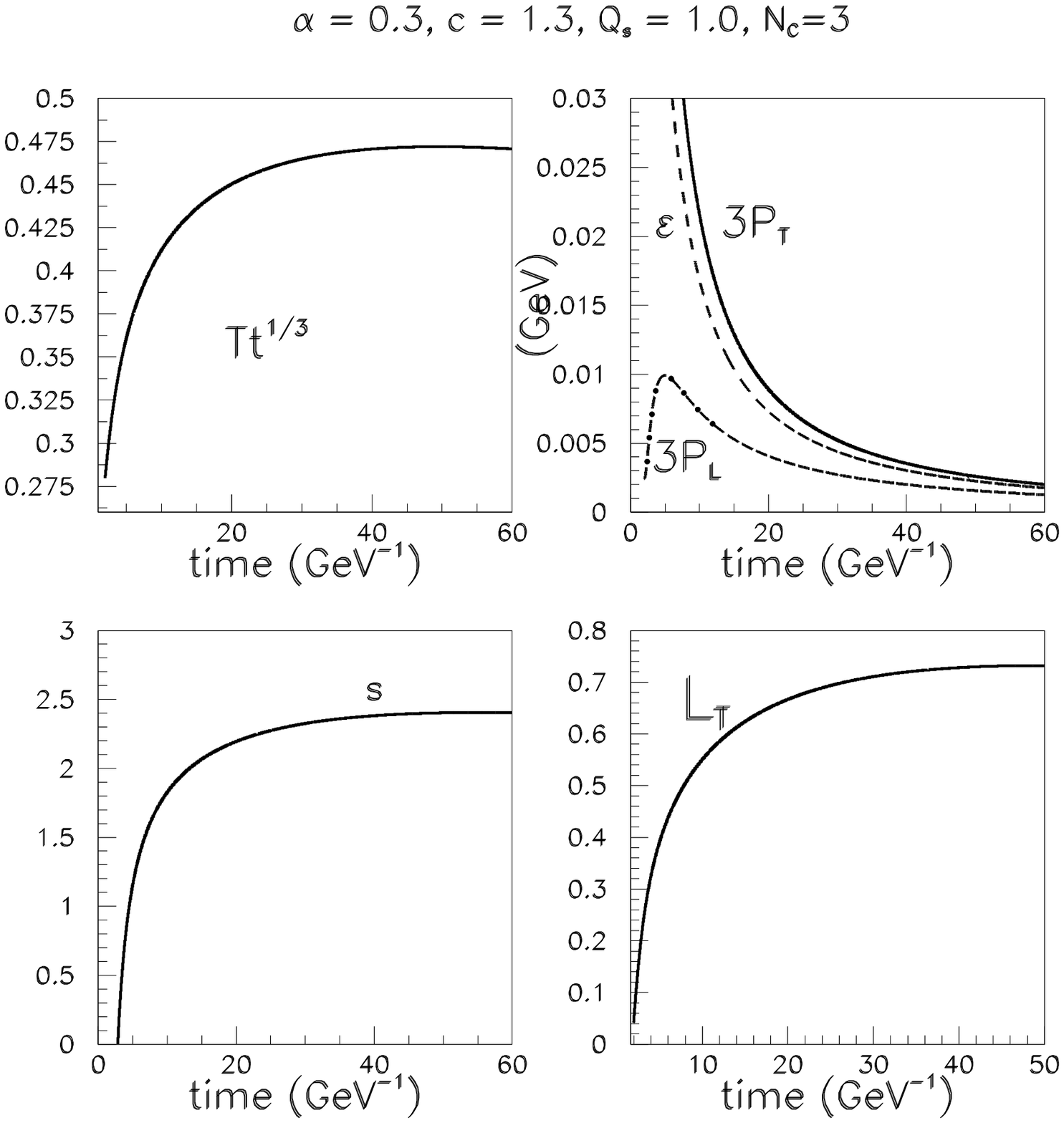,height=4.5in,width=4.in}
\caption{ $t^{1/3}$ times the energy
density per particle, the energy density $\epsilon$, and the longitudinal and transverse 
pressures, $P_L$ and $P_T$, the entropy density per
particle $s$ and $L$  plotted as functions of time.}
\label{PLOT2}
\end{figure}

Fig. \ref{PLOT2} shows, for $\alpha$, and $Q_s$ at RHIC energies, the energy density per
particle,  $\epsilon/n$, times $t^{1/3}$, the entropy density per particle and
logarithm $L$ versus time. As shown, $\epsilon t^{1/3}/n$ increases
monotonically and asymptotically approaches a constant, as expected at
kinetic equilibrium. The entropy density per particle $s$ also
monotonically increases and approaches a constant. This is also expected
since at equilibrium, $s = 3 - \mu/T$. Lastly, fig. \ref{PLOT2} shows $L$ (defined in
eq. \ref{DIV}) as a function of time. Since $\langle p\rangle =
\epsilon \propto T$ at kinetic equilibrium and $m_D^2 \propto 
T^2$, the argument of the logarithm in eq. (\ref{DIV})
approaches a constant at equilibrium. Fig. \ref{PLOT2} shows that $L$
is a constant at equilibrium. We define the time where kinetic
equilibrium begins to set in as the time $t_{eq}$, where $s$ and
$t^{1/3}\epsilon /n$  reach $10 \%$ of their asymptotic value.  
Having determined the time, $t_{eq}$, 
at equilibrium we determine the the temperature
$T_{eq}$, from (\ref{EQUIL}) and finally the chemical potential $\mu$.

In this way, for $\alpha_S=0.3$ and $N_c=3$,  we find the
equilibration time at RHIC ($Q_s = 1$ GeV), $t_{eq} = 3.24$
fm, the temperature $T_{eq} = 174.27$ MeV, and the chemical potential
$\mu = 157.86$ MeV. At LHC ($Q_s = 2\sim 3$ GeV), $t_{eq} = 1.42\sim 2.36$
fm, the temperature $T_{eq} = 320.72\sim471.69$ MeV, 
and $\mu=249.61\sim 457.72$ MeV.

\section{Conclusion}

In this paper, the Boltzmann equation for gluons was solved
numerically  for the single particle distributions $f(t,p)$. The time
evolution of $f(t,p)$ was followed to kinetic equilibrium. The
energy density $\epsilon$ and entropy density per particle $s$ were
determined. The results are expected to change if Bose 
enhancements are included in the Boltzmann equation\cite{DANIELWICZ,PRATT}and a
realistic initial single particle distribution, supplied by the calculation by
Krasnitz and Venugopalan\cite{ALEX_RAJU} is used as the initial
condition for $f$ in solving the Boltzmann equation.
As already mentioned, this equation only takes into account number conserving 
$2\longrightarrow 2$ processes\cite{MUELLER2}.  $2\longrightarrow 3$ processes may
have a large effect although how significant the effect may be still
remains to be determined.

\section{Acknowledgments}
This work was supported under DOE Contract No. DE--AC02--98CH10886 at BNL and 
DE-FG02-87ER40328 at the University of Minnesota.


\begin{thebibliography}{99}
\bibitem{MUELLER}
        A. H. Mueller. 
        Phys. Lett. {\bf B} 475, 220, (2000);
        A. H. Mueller. Nucl. Phys. {\bf B} 572, 227 (2000).
\bibitem{JEFF}
        J. Bjoraker and R. Venugopalan.
       (preprint) hep-ph/0008294.

\bibitem{ADRIAN}
       S.A. Bass, A. Dumitru,
       (preprint) nucl-th/0001033. 

\bibitem{BIRO}
        T. S. Biro, B. M\"uller and X. Wang,
        Phys. Lett. {283 \bf B}, 171 (1992);
        K. J. Eskola, B. M\"uller and X. Wang,
        Phys. Lett. {374 \bf B}, 21 (1996);
        L. Kadanoff and G. Baym, 
        ``Quantum Statistical Mechanics: Green's function methods in
         equilibrium and non-equilibrium problems'',
         Addison-Wesley Pub. Co., Advanced Book Program, (1989).

\bibitem{ALEX_RAJU}
        A. Krasnitz and R. Venugopalan,
        (preprint) hep-ph/0007108; 
        Phys. Rev. Lett. 84, 4309 (2000);
        Nucl. Phys. {\bf B} 557, 237 (1999);     

\bibitem{MV}L. McLerran and R. Venugopalan, {\em Phys. Rev.} {\bf D} 49, 
2233 (1994); {\em Phys. Rev.} {\bf D} 49, 3352 (1994); {\em Phys. Rev.}
 {\bf D} 50, 2225 (1994).

\bibitem{JKMW}J. Jalilian-Marian, A. Kovner, L. McLerran, and H. Weigert, 
{\em Phys. Rev.} {\bf D} 55, 5414 (1997). 

\bibitem{Kovchegov}Y.~V.~Kovchegov, {\em Phys. Rev.}  {\bf D54}, 5463 (1996).
   
\bibitem{BAYM}
         G. Baym, 
         Phys. Lett {138 \bf B} 19 (1984).
\bibitem{MUELLER2}  R. Baier, A.H. Mueller, D. Schiff  and D.T. Son,
        (preprint)  hep-ph/0009237. 
\bibitem{LIFSHITZ}
        E.M Lifshitz and L.P Pitaevskii,
        ``Physical Kinetics'', Pergamon Press, (1981).
\bibitem{BJORKEN}
         J.D. Bjorken,
         Phys. Rev. {\bf D} 27, 140  (1983).
\bibitem{DANIELWICZ}
       P. Danielewicz, Physica {\bf 100A}, 167 (1980).
\bibitem{PRATT}
       Scott Pratt and  Wolfgang Bauer, 
       Phys. Lett. {\bf B}329 413, (1994).

\end{thebibliography}
\end{document}